\documentclass[journal]{IEEEtran}

\usepackage{graphicx}
\usepackage{epsfig}
\usepackage{dcolumn}
\usepackage{bm}
\usepackage{amsmath}
\usepackage{amssymb}

\begin{document}
%
\title{Circuit elements with memory: {\it mem}ristors, {\it mem}capacitors and {\it mem}inductors}
%
%
%

\author{Massimiliano~Di~Ventra,
        Yuriy~V.~Pershin,
        and~Leon~O.~Chua,~\IEEEmembership{Fellow,~IEEE}
\thanks{M. Di Ventra is with the Department
of Physics, University of California, San Diego, La Jolla,
California 92093-0319 \newline e-mail: diventra@physics.ucsd.edu.}
\thanks{Yu. V. Pershin is with the Department of Physics
and Astronomy and USC Nanocenter, University of South Carolina,
Columbia, SC, 29208 \newline e-mail: pershin@physics.sc.edu.}
\thanks{L. O. Chua is with the Department of Electrical
Engineering and Computer Science, University of California,
Berkeley, California, 94720 \newline e-mail:
chua@eecs.berkeley.edu.}
\thanks{Manuscript received January XX, 2009; revised January YY, 2009.}}

%
%


\maketitle

\begin{abstract}
We extend the notion of memristive systems to capacitive and
inductive elements, namely capacitors and inductors whose
properties depend on the state and history of the system. All
these elements show pinched hysteretic loops in the two
constitutive variables that define them: current-voltage for the
{\it memristor}, charge-voltage for the {\it memcapacitor}, and
current-flux for the {\it meminductor}. We argue that these
devices are common at the nanoscale where the dynamical properties
of electrons and ions are likely to depend on the history of the
system, at least within certain time scales. These elements and
their combination in circuits open up new functionalities in
electronics and they are likely to find applications in
neuromorphic devices to simulate learning, adaptive and
spontaneous behavior.
\end{abstract}


%
\IEEEpeerreviewmaketitle

\section{Introduction}
%
%
%
%

\IEEEPARstart{C}{ircuits} elements that store information without
the need of a power source would represent a paradigm change in
electronics, allowing for low-power computation and storage. In
addition, if that information spans a continuous range of values
analog computation may replace the present digital one. Such a
concept is also likely to be at the origin of the workings of the
human brain and possibly of many other mechanisms in living
organisms so that such circuit elements may help us understand
adaptive and spontaneous behavior, or even learning.

One such circuit element is the memory-resistor (memristor for
short) which was postulated by Chua in 1971 by analyzing
mathematical relations between pairs of fundamental circuit
variables~\cite{litr2}. The memristor is characterized by a
relation between the charge and the flux, defined mathematically
as the time integral of the voltage, which need not have a
magnetic flux interpretation.  This relation can be generalized to
include any class of two-terminal devices (which are called
memristive systems) whose resistance depends on the internal state
of the system~\cite{litr3}.

Many systems belong to this class, including the thermistor
\cite{therm} (whose internal state depends on the temperature),
molecules whose resistance changes according to their atomic
configuration \cite{molec}, or spintronic devices whose resistance
varies according to their spin polarization \cite{litr4,litr5}.
Recently, memristive behavior and memory storage have been
reported in solid-state TiO$_2$ thin films \cite{HP1,HP2}, where
the change in resistance is realized by the ionic motion of oxygen
vacancies activated by current flow. Additionally, memristive
behavior has been demonstrated in VO$_2$ thin films, where the
memory mechanism is related to the insulator-to-metal transition
in these structures~\cite{Driscoll}. Finally, memristive behavior
has been identified by two of the present authors (MDV and YVP) as
a possible mechanism in the adaptive behavior of unicellular
organisms such as amoebas \cite{amoeba}.

All these examples show the ubiquitous nature of memristive
systems. In fact, it should not come as a surprise that many of
the above examples refer to nanoscale systems, whose resistance is
likely to depend on their state and dynamical history, at least
within (possibly very short) times scales dictated by the
fundamental state variables that control their operation
\cite{litr4}.

\begin{figure}[tb]
 \begin{center}
\includegraphics[angle=0,width=7.5cm]{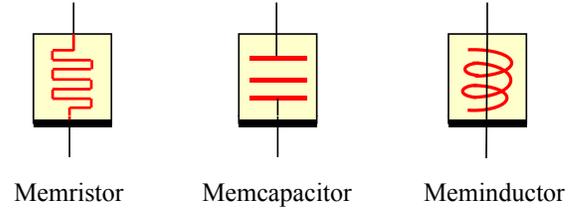}
\caption{\label{Figsymbols} (Color online) Symbols of the three
different devices defined in this work: memristor, memcapacitor,
and meminductor. In general, these devices are asymmetric and we
use the following convention when possible: when a positive
voltage is applied to the second (upper) terminal with respect to
the terminal denoted by the black thick line, the memory device
goes into a state of high resistance, capacitance or inductance,
respectively.}
 \end{center}
\end{figure}

In this paper we show that the above concept of memory device is
not necessarily limited to resistances but can in fact be
generalized to capacitative and inductive systems. Quite
generally, if $x$ denotes a set of $n$ state variables describing
the internal state of the system, $u(t)$ and $y(t)$ are any two
complementary constitutive variables~\cite{lochua} (i.e., current,
charge, voltage, or flux) denoting input and output of the system,
and $g$ is a generalized response, we can define a general class
of $n$th-order $u$-controlled memory devices as those described by
the following relations
\begin{eqnarray}
y(t)&=&g\left(x,u,t \right)u(t) \label{Geq1}\\ \dot{x}&=&f\left(
x,u,t\right) \label{Geq2}
\end{eqnarray}
where $f$ is a continuous $n$-dimensional vector function, and we
assume on physical grounds that, given an initial state $u(t=t_0)$
at time $t_0$, Eq.~(\ref{Geq2}) admits a unique
solution~\cite{prec1}.

{\it Memcapacitive} and {\it meminductive} systems are special
cases of Eqs.~(\ref{Geq1}) and~(\ref{Geq2}), where the two
constitutive variables that define them are charge and voltage for
the memcapacitance, and current and flux for the meminductance
(see Fig.~\ref{Figsymbols} for the symbols we use for these
devices). In this paper, we will discuss several properties of
these systems which make them unique compared to their
``standard'' definitions and will provide few examples. Combined
with the general class of memristive systems and/or standard
circuit elements these devices may lead to new functionalities in
electronics.

\section{Memristive systems}

For completeness, let us introduce first the notion of
memory-resistive systems \cite{litr3}. We also provide a simple
analytical example of memristive behavior which has been employed
in describing ``learning circuits''~\cite{amoeba}. For a broad
range of memristive system models, see, e.g.,
Ref.~\cite{Chuacourse}.

{\it Definition -} From Eqs.~(\ref{Geq1}) and~(\ref{Geq2}), an
$n$th-order current-controlled memristive system is described by
the equations

\begin{eqnarray}
V_M(t)&=&R\left(x,I,t \right)I(t) \label{eq1}\\
\dot{x}&=&f\left(x,I,t\right) \label{eq2}
\end{eqnarray}
with $x$ a vector representing $n$ internal {\it state variables},
$V_M(t)$ and $I(t)$ denote the voltage and current across the
device, and $R$ is a scalar, called the {\em memristance} (for
memory resistance) with the physical units of {\it Ohm}. The
equation for a charge-controlled memristor is a particular case of
Eqs. (\ref{eq1}) and (\ref{eq2}), when $R$ depends only on charge,
namely

\begin{equation}
V_M=R\left( q\left(t \right)\right)I, \label{eq3}
\end{equation}
with the charge related to the current via time derivative:
$I=dq/dt$. Note that the analytical equation describing the
TiO$_2$ device derived in the work by the Hewlett-Packard
group~\cite{HP1} has precisely this form, and it therefore
represents an {\it ideal} memristor.

We can also define an $n$th-order {\it voltage}-controlled
memristive system from the relations
\begin{eqnarray}
I(t)&=&G\left(x,V_M,t \right)V_M(t) \label{Condeq1}\\
\dot{x}&=&f\left( x,V_M,t\right) \label{Condeq2}
\end{eqnarray}
where we call $G$ the {\it memductance} (for memory conductance).

{\it Properties -} Several properties of memristive systems were
identified by Chua and Kang in their seminal paper \cite{litr3}.
Here, we highlight just a few of them. Considering devices with
$R\left( x,I,t\right)> 0$ in Eq. (\ref{eq1}), it has been proven
that these devices are passive \cite{litr3}. Another important
feature, following from the above inequality, is the so-called "no
energy discharge property" which is related to the fact that a
memristive system can not store energy, like a capacitor or an
inductor. As a manifestation of these characteristics, we notice
from Eq.~(\ref{eq1}) [or Eq.~(\ref{Condeq1})] that $V_M=0$
whenever $I=0$ (and vice versa). In addition, for a periodic
current input, the memristive system shows a ``pinched hysteretic
loop''~\cite{litr3}.

Moreover, a memristive system behaves as a linear resistor in the
limit of infinite frequency and as a non-linear resistor in the
limit of zero frequency provided that
$\dot{x}=f\left(x,I\right)=0$ in Eq.~(\ref{eq2}) (respectively
$\dot{x}=f\left( x,V_M\right)=0$ in Eq.~(\ref{Condeq2})) has a
steady-state solution. These two properties are easy to understand
on physical grounds. Irrespective of the underlying physical
mechanisms that define the state of the system, at very low
frequencies, the system has enough time to adjust its value of
resistance to a momentary value of the control parameter (either
current or voltage), so that the device behaves as a non-linear
resistor (from Eq.~(\ref{eq1}) or Eq.~(\ref{Condeq1})). On the
other hand, at very high frequency, there is not enough time for
any kind of resistance change during a period of oscillations of
the control parameter, so that the device operates as a usual
(linear) resistor.

Finally, it is worth pointing out that, when a current flows in a
resistor, at the microscopic level local resistivity dipoles form
due to scattering of carriers at interfaces~\cite{Landauer57}.
Namely, charges accumulate on one side of the resistor with
consequent depletion on the other side, with possible accumulation
and depletion over the whole spatial extent of the
resistor~\cite{DiVentra2008}. These dipoles develop {\em
dynamically} over length scales of the order of the screening
length~\cite{Sai}, namely within length scales over which
electrons can effectively screen the electric field of a local
charge. In other words, the self-consistent formation of local
resistivity dipoles takes some time and is generally accompanied
by some energy storage related to the capacitance of the system.
We thus expect that in nanoscale systems memristive behavior is
always accompanied to some extent by a (possibly very small)
capacitative behavior and as such the hysteresis loop may not
exactly cross the origin at frequencies comparable to the inverse
characteristic time of charge equilibration processes. This is,
for instance, found in the transverse voltage response of the spin
Hall effect in inhomogeneous semiconductor
systems~\cite{ourSHGpaper}.

\begin{figure}[tb]
 \begin{center}
\includegraphics[angle=0,width=7.5cm]{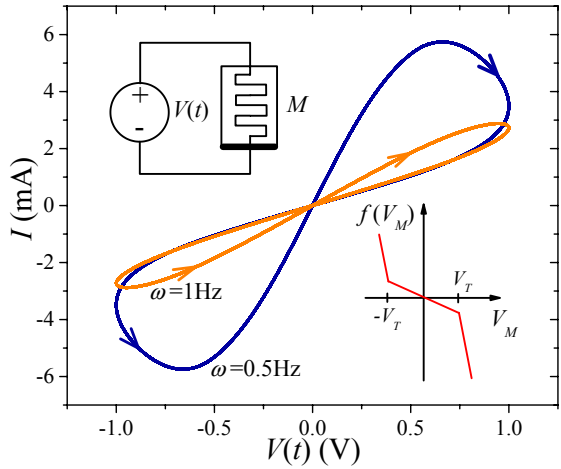}
\caption{\label{fig1} (Color online) Simulation of a
voltage-controlled memristive system. This plot was obtained by a
numerical solution of Eqs. (\ref{Condeq1}-\ref{Mmodel2}) for the
case of a memristive system $M$ connected directly to an
alternating (ac) voltage source $V(t)=V_0 \sin \left(2\pi \omega t
\right)$ as shown in the inset using the following parameter
values: $R_1=20\Omega$, $R_2=500\Omega$, $V_T=0.5$V,
$\alpha=-500$V/($\Omega\cdot$s), $\beta=2\alpha$, $V_0=1$V. In our
model, the rate of resistance change is defined by the function
$f(V_M)=-\left(\beta V_M+0.5\left( \alpha-\beta\right)\left[
|V_M+V_T|-|V_M-V_T| \right]\right)$ entering Eq. (\ref{Mmodel2})
and schematically shown in the inset for positive values of
$\alpha$ and $\beta$. At $|V_M|<V_T$ the slope of $f(V_M)$ is
determined by $\alpha$ and at $|V_M|>V_T$ it is determined by
$\beta$.}
 \end{center}
\end{figure}

{\it Examples -} Several systems have been found to satisfy the
above properties \cite{litr2,litr3,therm,HP1,HP2,litr4,Driscoll},
and many more are likely to be found which fit this
classification. In particular, memristive behavior is a property
of thermistors \cite{therm}, molecular systems \cite{molec},
spintronic devices \cite{litr4} and thin film nanostructures
\cite{HP1,HP2,Driscoll}, as well as numerous examples presented in
Ref.~\cite{Chuacourse}.

As an example, let us consider a model of a voltage-controlled
memristive system used by two of the present authors (MDV and YVP)
in Ref. \cite{amoeba}. In this model, the rate of system
resistance change ($R$ changes between two limiting values $R_1$
and $R_2$) is characterized by a constant $\alpha$ when $|V_M|\leq
V_T$ and $\beta$ when $|V_M|> V_T$, where $V_M$ is the voltage
applied to the memristor and $V_T$ is a threshold voltage
\cite{note1}. Mathematically, the response of this memristive
system is described by Eqs. (\ref{Condeq1}, \ref{Condeq2}) with

\begin{eqnarray}
G&=&x^{-1}, \\ \dot x&=&-\left(\beta V_M+0.5\left(
\alpha-\beta\right)\left[ |V_M+V_T|-|V_M-V_T| \right]\right)
\nonumber\\ & &\times \left[ \theta\left( V_M\right)\theta\left(
x-R_1\right) +\theta\left(- V_M\right)\theta\left(
R_2-x\right)\right] \label{Mmodel2},
\end{eqnarray}
where $\theta(\cdot)$ is the step function. The simulation
results~\cite{integro} depicted in Fig. \ref{fig1} clearly show a
pinched hysteresis loop and a hysteresis collapse with increasing
frequency $\omega$ of the alternating (ac) voltage source
$V(t)=V_0 \sin \left(2\pi \omega t \right)$.

We can now extend the above definitions to capacitances and
inductances. The memory devices that result share many
characteristics of memristive systems, but with a fundamental
difference: they store energy.

\section{Memcapacitative systems}

{\it Definition -} We define an $n$th-order voltage-controlled
memcapacitive system by the equations

\begin{eqnarray}
q(t)&=&C\left(x,V_C,t \right)V_C(t) \label{Ceq1} \\
\dot{x}&=&f\left( x,V_C,t\right) \label{Ceq2}
\end{eqnarray}
where $q(t)$ is the charge on the capacitor at time $t$, $V_C(t)$
is the corresponding voltage, and $C$ is the {\it memcapacitance}
(for memory capacitance) which depends on the state of the system.
Similarly, we can define an $n$th-order {\it charge}-controlled
memcapacitive system from the equations
\begin{eqnarray}
V_C(t)&=&C^{-1}\left(x,q,t \right)q(t) \label{CCeq1} \\
\dot{x}&=&f\left( x,q,t\right) \label{CCeq2}
\end{eqnarray}
where $C^{-1}$ is an inverse memcapacitance.

In addition, we define a subclass of the above devices, we call
voltage-controlled {\em memcapacitors}, when Eqs. (\ref{Ceq1})
and~(\ref{Ceq2}) reduce to

\begin{equation}
q(t)=C\left[\int\limits_{t_0}^tV_C\left( \tau \right)d\tau
\right]V_C(t) \label{VCMC}
\end{equation}
and a charge-controlled memcapacitor when Eqs. (\ref{CCeq1})
and~(\ref{CCeq2}) can be written as

\begin{equation}
V_C(t)=C^{-1}\left[\int\limits_{t_0}^tq\left( \tau \right)d\tau
\right]q(t), \label{CCMC}
\end{equation}
where in the above two equations, the lower integration limit
(initial moment of time) may be selected as $-\infty$, or 0 if
$\int_{-\infty}^0 V_C(\tau)d\tau=0$ and $\int_{-\infty}^0
q(\tau)d\tau=0$, respectively.

{\it Properties -} It follows from Eq.~(\ref{Ceq1}) that the
charge is zero whenever the voltage is zero. Note, however, that
in this case, $I=0$ does not imply $q=0$, and thus this device can
store energy. Therefore, the latter can be both added to and
removed from a memcapacitive system.

From a microscopic point of view a change in capacitance can occur
in two ways: {\it i)} due to a geometrical change of the system
(e.g., a variation in its structural shape), or {\it ii)} in the
quantum-mechanical properties of the carriers and bound charges of
the materials composing the capacitor (manifested, e.g., in a
history-dependent permittivity $\varepsilon$), or both.

In either case, inelastic (dissipative) effects may be involved in
changing the capacitance of the system upon application of the
external control parameter. These dissipative processes release
energy in the form of heating of the materials composing the
capacitor. However, this heat may not be simply accounted for as a
(time-dependent) resistance in series with the capacitor.

Similarly, there may be situations in which energy not from the
control parameter but from sources that control the equation of
motion for the state variable, Eq.~(\ref{Ceq2}), is needed to vary
the capacitance of the system (e.g., in the form of elastic energy
or via a power source that controls, say, the permittivity of the
system via a polarization field). This energy can then be released
in the circuit thus amplifying the current. Therefore,
Eqs.~(\ref{Ceq1}) and~(\ref{Ceq2}) for memcapacitive systems
postulated above could, in principle, describe both active and
passive devices.

\begin{figure}[tb]
 \begin{center}
\includegraphics[angle=0,width=6.5cm]{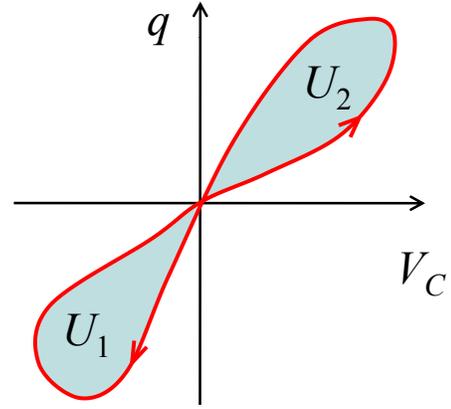}
\caption{(Color online) Schematics of a pinched hysteresis loop of
a memcapacitive system. The energy added to/removed from the
system ($\int V_C(q) dq$) is the area between the curve and the $q$
axis. The areas of shaded regions $U_1$ and $U_2$ give the amount
of added/removed energy in each half-period. The signs of $U_1$
and $U_2$ are determined by the direction on the loop. For the direction shown here, $U_2$ is positive and $U_1$ is negative. \label{fig2} }
 \end{center}
\end{figure}

However, starting from a fully discharged state, the amount of
removed energy from a {\it passive} memcapacitive system can not
exceed the amount of previously added energy. Mathematically, this
property can be written in the form
\begin{equation}
U_C=\int\limits_{t_0}^{t}V_C(\tau)I(\tau)d\tau\geq 0 \label{energ}
\end{equation}
under the condition that at $t=t_0$ no energy is stored in the
system. Eq. (\ref{energ}) should be valid for any form of the
control parameter, such as the voltage $V_C(t)$ applied to the
passive memcapacitive system. Moreover, in memcapacitive systems
with dissipative processes (such as heating), the equality sign in
Eq.~(\ref{energ}) for $t>t_0$ is not realized (assuming
$V_C(t)\neq 0$).

Fig. \ref{fig2} shows a schematic memcapacitive system hysteresis loop
passing through the origin. The shaded areas give the energy $U_i$ added
to/removed from the system. Note that
this energy is associated with some internal degree of freedom,
i.e., related to some physical elastic or inelastic processes that accompany the change in conductance.
The system is passive if
$U_1+U_2=0$, dissipative if $U_1+U_2>0$ and active if $U_1+U_2<0$.

\begin{figure}[tb]
 \begin{center}
\includegraphics[angle=0,width=7.5cm]{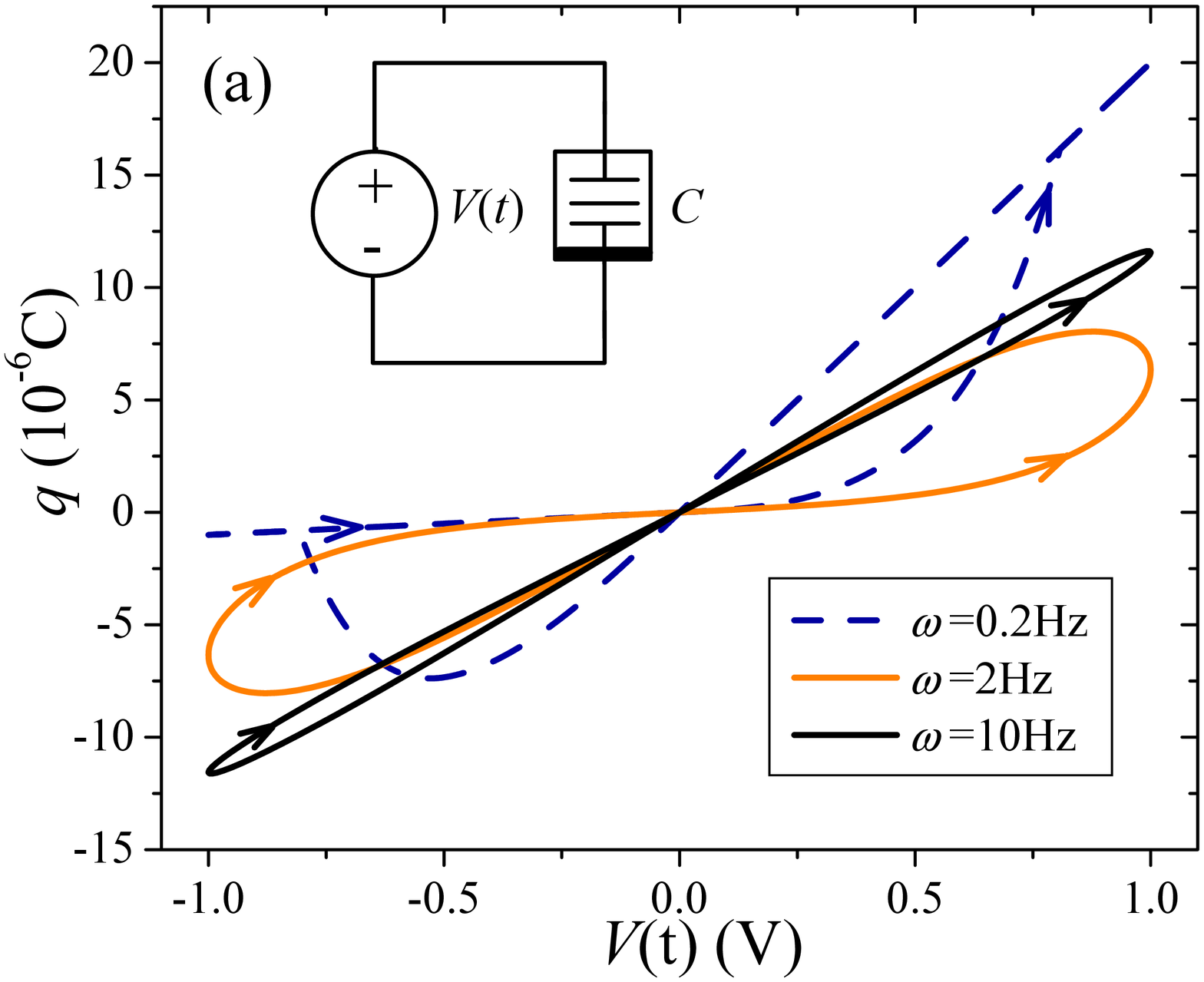}
\includegraphics[angle=0,width=7.5cm]{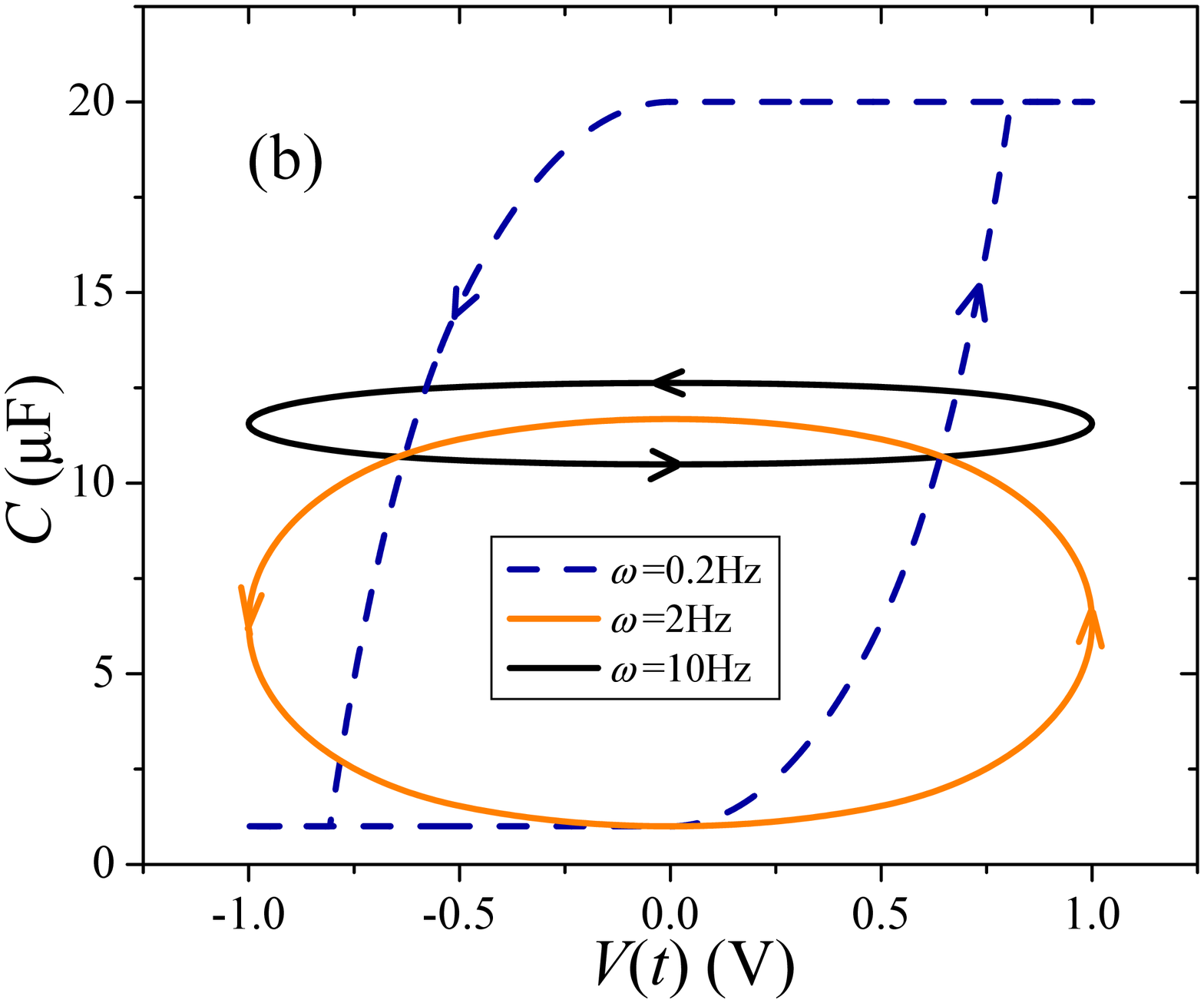}
\caption{(Color online) Simulation of a circuit (shown in the
inset of (a)) composed of a voltage-controlled memcapacitive
system and a resistance. The following parameter values were used
in this calculation: $C_1=1\mu$F, $C_2=20\mu$F, $V_T=0.5$V,
$\alpha=-50\mu$F/(V$\cdot$s), $\beta=2\alpha$, $V_0=1$V.
\label{fig3} }
 \end{center}
\end{figure}

Finally, since the state equation~(\ref{Ceq2}) has (by assumption)
only a unique solution at any given time $t\geq t_0$, then if
$V_C(t)$ is periodic with frequency $\omega$, namely $V_C(t)=V_0
\cos (2\pi \omega  t)$, then the $q-V_C$ curve is a simple loop
passing through the origin, namely there may be at most two values
of the charge $q$ for a given voltage $V_C$, if we consider a
voltage-controlled device (see Fig.~\ref{fig3}), or two values of
the voltage $V_C$ for a given charge $q$, for a charge-controlled
system. This loop is also anti-symmetric with respect to the
origin if, for the case of Eq. (\ref{Ceq1},\ref{Ceq2}),
$C\left(x,V_C,t \right)=C\left(x,-V_C,t \right)$ and
$f\left(x,V_C,t \right)=f\left(x,-V_C,t \right)$.

Like the case of memristive systems, a memcapacitive system
behaves as a linear capacitor in the limit of infinite frequency,
and as a non-linear capacitor in the limit of zero frequency,
assuming Eqs.~(\ref{Ceq2}) and~(\ref{CCeq2}) admit a steady-state
solution. The origin of this behavior rests again on the system's
ability to adjust to a slow change in bias (for low frequencies)
and the reverse: its inability to respond to extremely high
frequency oscillations.

{\it Examples -} There are instances in which the capacitance $C$
has been found to have a hysteresis-type dependence on the applied
voltage~\cite{Cexampl1,Cexampl2,Cexampl3}. These cases are related
to nanoscale capacitors in which interface traps \cite{Cexampl1}
or embedded nanocrystals \cite{Cexampl2,Cexampl3}  are responsible
for memory effects. Even though characteristic hysteresis
properties of these systems were not discussed in the framework of
the memcapacitive systems suggested in this paper, it is clear
that they belong to the present classification.

Below, we provide a model example of a memcapacitive system which
could be used to simulate experimental realizations of
memcapacitors. To be specific, let us consider a
voltage-controlled memcapacitive system according to
Eqs.~(\ref{Ceq1}) and~(\ref{Ceq2}). We focus on the electronic
circuit shown in the inset of Fig. \ref{fig3}. This scheme
consists of a memcapacitive system $C$ connected to a source of
ac-voltage $V(t)=V_0\sin(2 \pi \omega t)$. The capacitance $C$ is
chosen to vary in the range between its minimum and maximum values
($C_1$ to $C_2$) and changes according to the equations
\begin{eqnarray}
C&=&x \label{Cmodel1}\\ \dot x&=&-\left(\beta V_C+0.5\left(
\alpha-\beta\right)\left[ |V_C+V_T|-|V_C-V_T| \right]\right)
\nonumber\\ & &\times \left[ \theta\left( V_C\right)\theta\left(
x-C_1\right) +\theta\left(- V_C\right)\theta\left(
C_2-x\right)\right] \label{Cmodel2} .
\end{eqnarray}
Here, $V_C$ is the voltage drop on the memcapacitive system, $V_T$
is a threshold voltage, $\alpha$ is a variation rate for
$|V_C|<V_T$ and $\beta$ is a variation rate for $|V_C|>V_T$. The
interpretation of the coefficients $\alpha$ and $\beta$ is the
same as in Fig. \ref{fig1}. The response of the circuit is simply
described by the equation $V(t)=V_C=q/C$ where $C$ can be found
using Eqs. (\ref{Cmodel1},\ref{Cmodel2}). It is important to
remember that in modeling of more complex circuits the chain rule
of differentiation~\cite{note} should be used.

Fig. \ref{fig3} represents results of our numerical simulations.
Typical features discussed above such as capacitance hysteresis,
low-frequency and high-frequency behavior, and pinched hysteresis
loop can be clearly distinguished. We also checked the energy
conservation for the given circuit and found that for the present model the memcapacitive
system behaves as a passive device.

\section{Meminductive systems}

Let us now introduce the third class of memory devices. Let us
define the ``flux''
\begin{equation}
\phi(t)=\int\limits_{-\infty}^t V_L(t')dt',
\end{equation}
where $V_L(t)$ is the induced voltage on the inductor (equal to
minus the electromotive force).

{\it Definition -} We call an $n$th-order current-controlled
meminductive system one described by the equations

\begin{eqnarray}
\phi(t)&=&L\left(x,I,t \right)I(t) \label{Leq1} \\
\dot{x}&=&f\left(x,I,t\right) \label{Leq2}
\end{eqnarray}
where $L$ is the {\it meminductance}, and an $n$th-order {\it
flux}-controlled meminductive system the following

\begin{eqnarray}
I(t)&=&L^{-1}\left(x,\phi,t \right)\phi(t) \label{LLeq1} \\
\dot{x}&=&f\left( x,\phi,t\right) \label{LLeq2}
\end{eqnarray}
with $L^{-1}$ the inverse meminductance.

We call current-controlled {\it meminductors} a subclass of the
above devices, when Eqs. (\ref{Leq1}) and~(\ref{Leq2}) reduce to

\begin{equation}
\phi(t)=L\left[\int\limits_{t_0}^tI\left( \tau \right)d\tau
\right]I(t) \label{VCMC1}
\end{equation}
and a flux-controlled meminductor when Eqs. (\ref{LLeq1})
and~(\ref{LLeq2}) can be written as

\begin{equation}
I(t)=L^{-1}\left[\int\limits_{t_0}^t \phi\left( \tau \right)d\tau
\right]\phi(t). \label{CCMC1}
\end{equation}
In the above two equations, the lower integration limit may be
selected as $-\infty$, or 0 if $\int_{-\infty}^0 I(\tau)d\tau=0$
and $\int_{-\infty}^0 \phi(\tau)d\tau=0$, respectively.

{\it Properties -} For the sake of definiteness, we shall consider
the current-controlled meminductive systems. Taking the time
derivative of both sides of Eq. (\ref{Leq1}) yields

\begin{equation}
V_L=\frac{\textnormal{d}\phi}{\textnormal{d}t}=L\frac{\textnormal{d}I}{\textnormal{d}t}+
I\frac{\textnormal{d}L}{\textnormal{d}t}.
\end{equation}
The second term on the right-hand side of the above equation is an
additional contribution to the induced voltage due to a
time-dependent $L$. The energy stored in the current-controlled
meminductive system can be thus calculated as

\begin{equation}
U_L(t)=\int\limits_{t_0}^t
V_L(\tau)I(\tau)d\tau=\int\limits_{t_0}^t \left[
L\frac{\textnormal{d}I}{\textnormal{d}t}+
I\frac{\textnormal{d}L}{\textnormal{d}t}\right]I(\tau)d\tau.
\label{Lenerg}
\end{equation}

When $L$ is constant, we readily obtain the well-known expression
for the energy $U_L=LI^2/2$. Usually, this $U_L$ is interpreted as
the energy of the magnetic field generated by the current.
Similarly to the passivity criterion of memcapacitive systems (see
Eq. (\ref{energ})) Eq. (\ref{Lenerg}) provides the passivity
criterium of meminductive system if at $t=t_0$ the latter is in
its minimal energy state and $U_L(t)\geq 0$ at any time.


Finally, as in the case of memristive and memcapacitive systems,
it is expected that when the external control parameter is a
periodic function of time with frequency $\omega$, namely
$I(t)=I_0 \cos (2\pi \omega t)$, the $\phi-I$ curve is a simple
loop passing through the origin, namely there may be at most two
values of the flux $\phi$ for a given current $I$. This loop is
anti-symmetric with respect to the origin if, for the case of Eq.
(\ref{Leq1},\ref{Leq2}), $L\left(x,I,t \right)=L\left(x,-I,t
\right)$ and $f\left(x,I,t \right)=f\left(x,-I,t \right)$.

Also, the magnetic permeability can adjust to slow periodic
variations of the current, while it cannot at high frequencies, so
that this device behaves as a non-linear inductor at low
frequencies (assuming Eqs.~(\ref{Leq2}) and~(\ref{LLeq2}) admit a
steady-state solution) and a linear inductor at high frequencies.

{\it Examples -} In electronics, inductors are primarily of a
solenoid type consisting of a coil of conducting material wrapped
around a core. The inductance of a solenoid is proportional to the
relative permeability $\mu_r$ of the material within the solenoid
and also depends on the geometrical parameters of the system. The
simplest way to introduce a memory effects in such a system is to
use the core material whose response to the applied magnetic field
depends on its history. As an example, we can think about
ferromagnetic materials exhibiting a magnetic hysteresis such as
iron. In fact, an electronic circuit with such inductor was
recently analyzed \cite{Matsuo}. Another way to introduce memory
effects is by varying the inductor's shape.

A circuit model for a meminductive system can be formulated
similarly to the models of memristive and memcapacitive systems
discussed above. In particular, we can imagine a structure whose
inductance increases up to a certain value when current flows in
one direction and decreases down to another value if the current
flow direction is reversed. This is a straightforward calculation
which does not add much to the analysis of this problem.
Therefore, we do not report it here.

\section{Conclusions}

We have extended the notion of memory devices to both capacitative
and inductive systems. These devices have specific properties that
appear most strikingly as a pinched hysteretic loop in the two
constitutive variables that define them: current-voltage for
memristive systems, charge-voltage for memcapacitive systems, and
current-flux for the meminductive systems. Many systems belong to
these classifications, especially those of nanoscale dimensions.
Indeed, with advances in the miniaturization of devices, these
concepts are likely to become more relevant since at the nanoscale
the dynamical properties of electrons and ions may strongly depend
on the history of the system, at least within certain time scales~\cite{DiVentra2008}.

Apart from the obvious use of these devices in non-volatile
memories, several applications can be already envisioned for these
systems, especially in neuromorphic devices to simulate learning,
adaptive and spontaneous behavior. For instance, the
identification of memristive behavior in primitive organisms such
as amoebas~\cite{amoeba}, opens up the possibility to relate
physiological processes that occur in cells with the theory of
memory devices presented here. Along similar lines, one could
envision simple models that identify memory mechanisms in neurons
and use these memory devices to build such models in the
laboratory.

Therefore, due to their versatility (including analog
functionalities) the combined operations of these memory devices
in electronic circuits is still largely unexplored, and we hope
our work will motivate experimental and theoretical investigations
in this direction.

\section*{Acknowledgment}

The authors are indebted to B. Mouttet for pointing out
Ref.~\cite{Cexampl2}. This work has been partially funded by the
NSF grant No. DMR-0802830.

\ifCLASSOPTIONcaptionsoff
  \newpage
\fi


\begin{thebibliography}{100}
\bibitem{note1} This model was inspired by an
activation-type resistance state found in nanostructure memristors
\cite{HP2}.
\bibitem{integro} All numerical calculations presented in this paper have been performed using a fourth-order Rugge-Kutta integration method.
\bibitem{note} In the case of a memory capacitor, since the
capacitance $C$ is time-dependent the well-known relation
$\textnormal{d}q=C\textnormal{d}V_C$ generalizes to
$\textnormal{d}q=C\textnormal{d}V_C+V_C\textnormal{d}C$.
\end{thebibliography}
\end{document}